\documentstyle[11pt, aaspp4,epsf,rotate]{article}

\def\atan{{\rm arctan}}

\lefthead{B\"ottcher \& Liang}
\righthead{Alternating phase lags}

\begin{document}

\title{A model for the alternating phase lags associated with QPOs
in X-ray binaries}
\author{M.  B\"ottcher\footnote{Chandra Fellow}\altaffilmark{2}, 
E. P. Liang\altaffilmark{2}}
\altaffiltext{2}{Rice University, Space Physics and Astronomy Department, MS 108\\
6100 S. Main Street, Houston, TX 77005 -- 1892, USA}

\begin{abstract}
We present a theoretical model for the alternating phase lags
associated with QPO fundamental and harmonic frequencies observed
in some Galactic black-hole candidates. We assume that the accretion
flow exhibits a transition from an outer cool, optically thick 
accretion disk to a hot, inner advection-dominated accretion flow 
(ADAF), and that the QPOs are related to small-scale oscillations 
of the accretion rate and the transition radius. We present an 
analytical estimate of the expected phase lags at the fundamental 
and first harmonic frequencies of the QPOs and perform detailed 
time-dependent Monte-Carlo simualtions of the radiation transport 
in the oscillating ADAF / cool disk system. We find that this model 
is well suited to reproduce alternating phase lags between the 
fundamental and the first harmonic. It also naturally explains 
the trend observed in GRS~1915+105 that, as the soft X-ray luminosity 
increases, the QPO frequency increases and the phase lag associated 
with the QPO fundamental frequency changes sign from positive to 
negative. The relation between the disk temperature and the QPO 
frequency observed in GRS~1915+105 is consistent with a secular 
instability modulating the disk evaporation at the transition 
radius.
\end{abstract}


\section{Introduction}

The steadily increasing amount of high-quality, high time-resolution
X-ray data from Galactic X-ray binaries, has stimulated vital interest 
in the characteristics of the rapid variability of these objects. The
X-ray emission from both Galactic black-hole candidates (GBHCs) and
accreting neutron stars is known to vary on a wide range of time scales,
sometimes showing quasi-periodic oscillations (QPOs) (for reviews see,
e.g., \cite{vdk95,cui99a}). In GBHCs, such QPOs have been detected at
frequencies ranging from several mHz to $\lesssim 100$~Hz. These QPOs
are most notable when the sources are in the low-hard or in the rare
very high state. Recently, correlations between the amplitudes and 
centroid frequencies of several types of QPOs with the spectral
characteristics of the source have been found in some objects
(\cite{rutledge99}, Markwardt, Swank, \& Taam \markcite{mst99}1999, 
\cite{sobczak99}, Muno, Morgan \& Remillard \markcite{mmr99}1999).
A general pattern emerging from these analyses is that the frequency 
of those types of QPOs which do show a correlation with spectral
properties, seems to increase with both the power-law and the disk
black-body flux from the source. 

A surprising property of some types of QPOs has recently been found
in the 67 mHz QPO of GRS~1915+105 (\cite{cui99b}), the 0.5 -- 10~Hz
QPO of the same source (\cite{reig00}, \cite{lin00}), and in the 0.3 
-- 3~Hz QPO in XTE~J1550-564 (Cui, Zhang, \& Chen \markcite{cui00}2000): 
While the phase lag at the QPO fundamental frequency is negative, the 
phase lag associated with the first harmonic was positive. In the case 
of the low-frequency QPO of GRS~1915+105, even three harmonics were 
detected, and the phase lags were found to alternate between subsequent 
harmonics (\cite{cui99b}). The phase lag associated with the 0.5 -- 10~Hz
QPO of GRS~1915+105 was found to change sign from positive to negative
as the QPO frequency increases above $\sim 2.5$~Hz (\cite{reig00},
\cite{lin00}). 

These peculiar patterns are apparently completely counter-intuitive
in the light of currently discussed models for the hard phase lags 
in X-ray binaries. Models proposed to explain the hard phase lags
are either based to the energy-dependent photon escape time in
Compton upscattering scenarios for the production of hard X-rays
(Kazanas, Hua \& Titarchuk \markcite{kht97}1997, B\"ottcher \& 
Liang \markcite{bl98}1998), or due to intrinsic spectral hardening
during X-ray flares, e.g. due to decreasing Compton cooling in
active regions pushed away from an underlying accretion disk in
a patchy-corona model (\cite{pf99}) or due to density perturbations 
drifting inward through an ADAF toward the event horizon 
(B\"ottcher \& Liang \markcite{bl99}1999). These models dealt
only with the continuum variability and did not consider the
effects of QPOs. 

In this paper we will explore the response of a two-phase accretion
flow, consisting of an outer, cool, optically thick accretion disk
and an inner, hot ADAF (\cite{ny94}, \cite{abramowicz95}, \cite{cal95}), 
to a periodically varying soft photon input from the cool disk
(\cite{lb00}). A two-phase accretion flow with an inner ADAF 
has been found to produce good fits to the photon spectra of, 
e.g., several Galactic X-ray binaries (e.g., Narayan, McClintock 
\& Yi \markcite{nmy96}1996, \cite{hameury97}, \cite{esin98}), 
low-luminosity AGN (\cite{quataert99}) and giant elliptical 
galaxies (\cite{fr95}, \cite{df97}, \cite{dimatteo00}). In 
\S \ref{analytical}, we describe the basic model setup according 
to this two-phase accretion flow, and derive an analytical estimate 
for the expected phase lags associated with the QPO and the first 
harmonic applicable in some simplified cases. A short description 
of the Monte-Carlo simulations used to solve the time-dependent 
radiation transport problem follows in \S \ref{mc}. In \S \ref{grs1915}
we describe a series of simulations designed specifically to explain
the peculiar phase lag behavior associated with the 0.5 -- 10~Hz
QPO in GRS~1915+105. We summarize in \S \ref{summary}.

\section{\label{analytical}Model setup and analytical estimates}

The basic model setup, as motivated in the introduction, assumes 
that the inner portion of the accretion flow is described by an
ADAF, where the density profile is approximately given by a free-fall
profile, $n_{\rm e} (r) \propto r^{-3/2}$, and the electron temperature 
is close to the virial temperature and thus scales as $T_{\rm e} (r) 
\propto r^{-1}$. We emphasize, however, that other hot two-temperature 
inner disk models would give similar results. The ADAF exists out 
to a transition radius $R_{\rm tr}$, beyond which the flow is 
organized in a standard optically thick, geometrically thin 
Shakura-Sunyaev disk (\cite{ss73}). The transition radius is 
typically expected to be several~$100 \, R_s \lesssim R_{\rm tr} \lesssim 
10^4 \, R_s$ (\cite{honma96,manmoto00,meyer00}), where $R_s$ is the
Schwarzschild radius. Outside this transition radius, the disk
temperature scales as $T_{\rm D} (r) \propto r^{-3/4}$. With the free-fall
density profile given above, the radial Thomson depth of the corona 
may be written as $\tau_{\rm T}^{\rm ADAF} = 0.71 \, \dot M_{17} / (m \, 
r_i^{1/2})$, where $\dot M_{17}$ is the accretion rate through the ADAF
in units of $10^{17}$~g~s$^{-1}$, $m$ is the mass of the central compact 
object in units of solar masses, and $r_i$ is the inner edge of the ADAF
in units of Schwarzschild radii.

The basic assumption of our baseline model is that the observed QPOs in 
the X-ray variability of Galactic black-hole candidates are related to
small-scale oscillations of the transition radius. Such oscillations
in $r_{\rm tr}$ may be caused by variations of the accretion rate, but
we defer a detailed study of possible mechanisms driving the oscillations
to a later paper. We thus assume that, as a function of time $t$, 
the transition radius oscillates as $R_{\rm tr}(t) = r_{\rm tr} \, 
(1 + a_r \, \sin[\omega t])$, where $a_r \ll 1$ is the amplitude 
of the oscillation and $\omega = 2 \pi f_{\rm QPO}$ the QPO frequency. 
Denoting $\xi \equiv 1 + a_r \, \sin(\omega t)$, the time-dependent disk 
flux may then be estimated as $F_{\rm D} (t) \propto [R_{\rm tr} (t)]^2 \, 
[T_d (t)]^4 \propto (r_{\rm tr} \, \xi)^{-1}$. In the following, 
we are focusing on an analytical description of the X-ray signals 
at the QPO fundamental and first harmonic frequencies. Thus, we will 
expand all expressions up to the 2nd order in $a_r \sin(\omega t)$.
For the disk flux, this yields

\begin{equation}
F_{\rm D} (t) \approx F_{\rm D, 0} \, \left[ 1 + {a_r^2 \over 2} - a_r 
\, \sin(\omega \, t) - {a_r^2 \over 2} \, \cos(2 \omega \, t) \right].
\end{equation}
The disk spectrum is approximately a blackbody spectrum at the
temperature of the disk at the transition radius, and for the
purpose of a simple analytical estimate, we assume that it is 
monochromatic at a disk photon energy $E_{\rm D} (t) \propto 
T_{\rm D}(r_{\rm tr}) \propto (r_{\rm tr} \, \xi)^{-3/4}$. For 
any observed photon energy $E$, we define the ratio 

\begin{equation}
\epsilon (t) \equiv {E / E_{\rm D} (t)} \equiv \epsilon_0 \, \xi^{3/4}.
\end{equation}

Now, at any given time, a fraction $f_c \approx {1 \over 2} \, 
\left(1 - {\pi \over 4} \right)$ of the disk radiation will intercept
the quasi-spherical inner ADAF-like corona and serve as soft seed
photons for Compton upscattering, producing the time-variable hard 
X-ray emission. Note that additional hard X-ray emission will be
produced by Comptonization of synchrotron and bremsstrahlung 
photons produced in the inner portions of the ADAF. However, 
since we do not assume significant changes of the structure 
of the inner ADAF in the course of the small-scale oscillations 
of the transition radius, this internal synchrotron and bremsstrahlung
emission will constitute a quasi-DC flux component which does not 
contribute significantly to the variability properties of the source. 
Being mainly interested in the quasi-periodic variability, we do not 
consider this emission component here. The time-dependent Comptonization 
response function to the oscillating disk flux in this geometry can be 
parametrized as

\begin{equation}
h (t, \tau, \epsilon_0) = \Theta (u) \, \left\lbrace A \, 
u^{\alpha - 1} \, e^{-u / \beta} + B \, u^{\kappa - 1} \, 
\Theta \left( {2 \, R_{\rm tr} (\tau) \over c} - u\right) \right\rbrace
\label{transferfunction}
\end{equation}
(\cite{bl98}), where now $t$ is the observing time, $\tau$ is the
time of soft photon emission at the transition radius, $u \equiv
t - \tau$, $\Theta$ is the Heaviside function, $A$ and $B$ are 
normalization factors which are generally energy and time dependent. 
In the general case, the indices $\alpha$ and $\kappa$, and the 
time ``constant'' $\beta$ also depend on photon energy and time
($\beta$ depends on time through the time-dependence of the extent
of the corona and the disk photon energy). 

The observable energy flux at time $t$ at photon energy $E$ is then
given by

\begin{equation}
F (E, t) = f_d \, F_{\rm D} (E, t) + f_c \int\limits_{- \infty}^{t} d\tau
\int\limits_0^{\infty} dE_{\rm D} \; F_{\rm D} (E_{\rm D}, \tau) \; h(t, \tau, \epsilon_0),
\label{total_em}
\end{equation}
where $f_d \equiv 1 - f_c$. For the purpose of the analytical estimate,
we now focus on two regimes for the observed photon energies. At low
energies, close to the peak energy of the disk emission, the observed 
light curve will be dominated by the direct disk emission (the first 
term on the r.h.s. of Eq. \ref{total_em}) and the contribution from 
reflection in the outer regions of the ADAF, proportional to $B 
\approx c / (2 R_{\rm tr} [\tau])$. For an ADAF-like density and 
temperature profile, we may set $\kappa = 1$, i. e. the reflection 
light curve is flat over the light crossing time through the corona. 
The multiple-scattering term, proportional to $A$ in Eq. 
\ref{transferfunction}, may be neglected at low photon energies.

The observed signal at high photon energies, $E \gg E_{\rm D}$ will 
be dominated by the multiple-scattering term, i. e. we may set 
$B = 0$ for high-energy photons. Since multiple Compton scattering 
results in a power-law spectrum, we parametrize the energy
dependence of the normalization of this term as $A = A_0 \, 
\epsilon_0^{-\gamma} \, x^{- 3 \gamma / 4}$, where $\gamma$ is
the photon spectral index of the Comptonized spectrum, and $x = 1 +
\sin(\omega \tau)$. The parameter $\alpha$ is generally energy
dependent. However, we find that a constant value of $\alpha = 10$ 
yields a reasonably good description of the multiple-scattering
light curve for an ADAF-like temperature and density profile.
To the same degree of approximation, we consider $\beta$ as constant
for a given energy channel, and note its scaling as $\beta \propto 
r_{\rm tr}$. Furthermore, $\beta$ increases with photon energy. Now, 
using the $\delta$ function assumption for the disk spectrum in 
the integral in Eq. \ref{total_em}, the observed light curve reduces 
to

\begin{equation}
F (E, t) \approx f_d \, F_{\rm D} (E, t) + f_c \, F_{D, 0} 
\int\limits_0^{\infty} du \; \left\lbrace A_0 \, 
\epsilon_0^{-\gamma} \, x^{- \delta} u^{\alpha - 1} 
e^{-u / \beta} \; + \; x^{-2} B_0 (\epsilon_0)
\, \Theta \left( {2 \, r_{\rm tr} \over c} - u \right) 
\right\rbrace,
\end{equation}
where $\delta \equiv (3 \gamma / 4) + 1$. 

Expanding all terms up to the first harmonic and writing

\begin{equation}
{F (E, t) \over F_{D, 0}} = \eta_0 + \eta_1 \sin(\omega t) + 
\eta_2 \cos(\omega t) + \eta_3 \sin(2 \omega \, t) + 
\eta_4 \cos(2 \omega \, t)
\end{equation}
we find

\begin{equation}
\eta_0 = f_d \, \left( 1 + {a_r^2 \over 2} \right) + f_c \, {B
\over \omega} \, \phi_r \, \left( 1 + {3 \over 2} a_r^2 \right)
+ f_c A_0 \epsilon_0^{- \gamma} \left( \beta^{\alpha} \Gamma[\alpha]
+ {\delta [\delta + 1] \over 4} a_r^2 \, [C_2 (\alpha) + S_2 (\alpha)]
\right),
\label{eta0}
\end{equation}
\begin{equation}
\eta_1 = - f_d \, a_r - 2 \, f_c {B \over \omega} a_r \sin\phi_r
- f_c A_0 \epsilon_0^{- \gamma} \delta a_r \, C_1 (\alpha),
\label{eta1}
\end{equation}
\begin{equation}
\eta_2 = 2 \, f_c {B \over \omega} a_r \, (1 - \cos\phi_r) 
+ f_c A_0 \epsilon_0^{-\gamma} \delta a_r S_1 (\alpha),
\label{eta2}
\end{equation}
\begin{equation}
\eta_3 = - {3 \over 2} f_c {B \over \omega} a_r^2 \sin^2 \phi_r
- {1 \over 2} f_c A_0 \epsilon_0^{-\gamma} \delta (\delta + 1)
\, a_r^2 SC (\alpha),
\label{eta3}
\end{equation}
\begin{equation}
\eta_4 = - {1 \over 2} f_d \, a_r^2 - {3 \over 2} f_c {B \over \omega}
a_r^2 \, \sin\phi_r \, \cos\phi_r + {1 \over 4} f_c A_0 \epsilon_0^{-\gamma}
\delta (\delta + 1) \, a_r^2 (S_2 [\alpha] - C_2 [\alpha]),
\label{eta4}
\end{equation}
where $\phi_r \equiv {2 \omega \over c} \, r_{\rm tr}$, and we have defined
the integrals

\begin{equation}
S_1 (\alpha) \equiv \int\limits_0^{\infty} du \; u^{\alpha - 1} 
e^{-u / \beta} \sin(\omega u) = \beta^{\alpha} \Gamma (\alpha)
{\sin (\alpha \, {\rm arctan} [\beta\omega]) \over (1 + [\beta\omega]^2
)^{\alpha / 2}},
\end{equation} 
\begin{equation}
C_1 (\alpha) \equiv \int\limits_0^{\infty} du \; u^{\alpha - 1} 
e^{-u / \beta} \cos(\omega u) = \beta^{\alpha} \Gamma (\alpha)
{\cos (\alpha \, {\rm arctan} [\beta\omega]) \over (1 + [\beta\omega]^2
)^{\alpha / 2}},
\end{equation} 
\begin{equation}
S_2 (\alpha) \equiv \int\limits_0^{\infty} du \; u^{\alpha - 1} 
e^{-u / \beta} \sin^2(\omega u) = {\beta^{\alpha} \Gamma (\alpha)
\over 2} \left\lbrace 1 - {\cos (\alpha \, {\rm arctan} [2 \, 
\beta\omega]) \over (1 + [2 \, \beta\omega]^2)^{\alpha / 2}}
\right\rbrace,
\end{equation}
\begin{equation}
C_2 (\alpha) \equiv \int\limits_0^{\infty} du \; u^{\alpha - 1} 
e^{-u / \beta} \cos^2(\omega u) = {\beta^{\alpha} \Gamma (\alpha)
\over 2} \left\lbrace 1 + {\cos (\alpha \, {\rm arctan} [2 \, 
\beta\omega]) \over (1 + [2 \, \beta\omega]^2)^{\alpha / 2}}
\right\rbrace,
\end{equation}
\begin{equation}
SC (\alpha) \equiv \int\limits_0^{\infty} du \; u^{\alpha - 1} 
e^{-u / \beta} \sin(\omega u) \cos(\omega u) = {\beta^{\alpha} 
\Gamma (\alpha) \over 2} {\sin (\alpha \, {\rm arctan} [2 \, 
\beta\omega]) \over (1 + [2 \, \beta\omega]^2)^{\alpha / 2}}.
\end{equation}
Defining the phases, relative to the $R_{\rm tr}$ oscillation,
of the signal in a given energy band at the QPO frequency and 
the first harmonic, respectively, by

\begin{equation}
F(t) = \eta_0 + \xi_1 \sin (\omega t + \Delta_Q) + \xi_2 \sin
(2 \omega \, t + \Delta_H)
\label{Delta}
\end{equation}
with $\tan \Delta_Q = \eta_2 / \eta_1$ and $\tan \Delta_H = \eta_4
/ \eta_3$, we find for a low-energy channel (with mean energy close
to the peak energy of the disk emission spectrum):

\begin{equation}
\tan\Delta_Q ({\rm LE}) \approx - {2 \, f_c {B \over \omega} \,
(1 - \cos\phi_r) \over 2 \, f_c {B \over \omega} \, \sin\phi_r
+ f_d },
\label{tan_Q_le}
\end{equation}
\begin{equation}
\tan\Delta_H ({\rm LE}) \approx {f_d +{3 \over 2} \, f_c
{B \over \omega} \, \sin(2 \phi_r) \over {3 \over 2} \, f_c
{B \over \omega} \, (1 - \cos[2 \phi_r]) }.
\label{tan_H_le}
\end{equation}
Assuming that the direct disk emission ($\propto f_d$) is strongly 
dominating at this energy, and that both $\eta_3$ and $\eta_4$ are 
negative in this case, we have $\Delta_Q ({\rm LE}) \approx 0$, and
$\Delta_H ({\rm LE}) \approx - {\pi/2}$. 

For high photon energies, dominated by multiple Compton scattering,
we find

\begin{equation}
\Delta_Q ({\rm HE}) = - \alpha \, \atan (\beta\omega) + k_0 \pi,
\label{tan_Q_he}
\end{equation}
\begin{equation}
\Delta_H ({\rm HE}) = \alpha \, \atan (2 \, \beta\omega) + k_1 \pi,
\label{tan_H_he}
\end{equation}
where $k_i \in (0, -1, 1)$ are determined by the signs of $\eta_1$,
\dots, $\eta_4$. With the definition of the phases $\Delta$ in Eq.
\ref{Delta}, the phase lags are given by $\Delta\phi = \Delta ({\rm
LE}) - \Delta ({\rm HE})$. Throughout this paper we adopt the 
convention that positive phase and time lags correspond to hard 
photons lagging the soft ones.

Intuitively, phase and time lags between the disk-radiation dominated 
low-energy photons and the Compton-upscattering dominated high-energy
photons are determined by the ratio between the QPO period $T_{\rm QPO}$ 
and the time required for soft photons to reach the inner ADAF region 
and be Compton upscattered to hard X-ray energies. If this light-travel 
and diffusion time is {\it longer than half a QPO period, but less than 
the QPO period itself}, the phase lag at the QPO fundamental frequency 
--- which physically is still a hard lag --- appears as a {\it soft lag }
due to the periodicity of the light curves. The same argument holds 
for the $n^{th}$ harmonic of the QPO, where the period now corresponds 
to $1/(n + 1)$ of the QPO fundamental period.

For a very large transition radius, the direct disk emission may not 
contribute significantly to the X-ray flux. In that case, the phase 
lag between two energy channels will be dominated by the difference 
in diffusion time scales $\beta$, and we expect

\begin{equation}
\Delta\phi_{\rm QPO} \approx \alpha (\atan [\beta_j \omega] 
- \atan [\beta_i \omega]),
\end{equation}
\begin{equation}
\Delta\phi_{\rm 1. \; harm.} \approx \alpha (\atan [2 \beta_i \omega]
- \atan [2 \beta_j \omega]).
\end{equation}
for two energy channels $E_i < E_j$.

\section{\label{mc}Monte Carlo simulations}

The estimates derived in the previous section were based on several
major simplifications in order to keep the problem analytically
trackable. In order to test the validity of our approximations and
to provide results under more realistic assumptions, we have simulated
the radiation transfer in our oscillating ADAF/disk model system, 
using our time-dependent Monte-Carlo Comptonization code. For a 
detailed description of the code and its capabilities, see 
B\"ottcher \& Liang (\markcite{bl98}1998, \markcite{bl99}1999). 

In our simulations, we approximate the accretion disk emission by
a blackbody spectrum with the temperature of the disk inner edge
at any given time, and its time-dependent luminosity determined 
as described in the previous section. A fraction $f_c$ of the time
dependent disk emission enters the spherical Comptonizing region at 
its outer boundary and serves as seed photon field for Comptonization. 
The ADAF is characterized by an $r^{-3/2}$ density and $r^{-1}$ temperature 
structure, and the electron temperature is normalized to $kT_e (R_s) 
= 500$~keV, i. e. the electrons become relativistic at the event
horizon. The event horizon is treated as an absorbing inner boundary.
The corona has the specified radial Thomson depth of $\tau_{\rm T}^{\rm ADAF}$ 
when the ADAF/disk transition is located at $r_{\rm tr}$. As $r_{\rm tr}$ 
oscillates, we leave the density and temperature structure of the 
ADAF inside $R_{\rm tr} (t)$ unchanged and set the coronal electron 
density equal to zero outside $R_{\rm tr} (t)$. 

The resulting light curves, generally consisting of the sum of direct 
disk emission and the Comptonized emission from the corona, are sampled
in 5 photon energy bins and over 512 time steps of $\Delta t \lesssim 
0.05 \, T_{\rm QPO}$. For several test cases, we have run identical 
problem simulations with different time steps, in order to verify that 
our results are independent of the time step chosen. The energy-dependent
light curve are then Fourier transformed, using an FFT algorithm
(\cite{press92}), and the power density spectra and phase lags are 
calculated.

\begin{figure}
\rotate[r]{
\epsfysize=12cm
\epsffile[150 0 550 500]{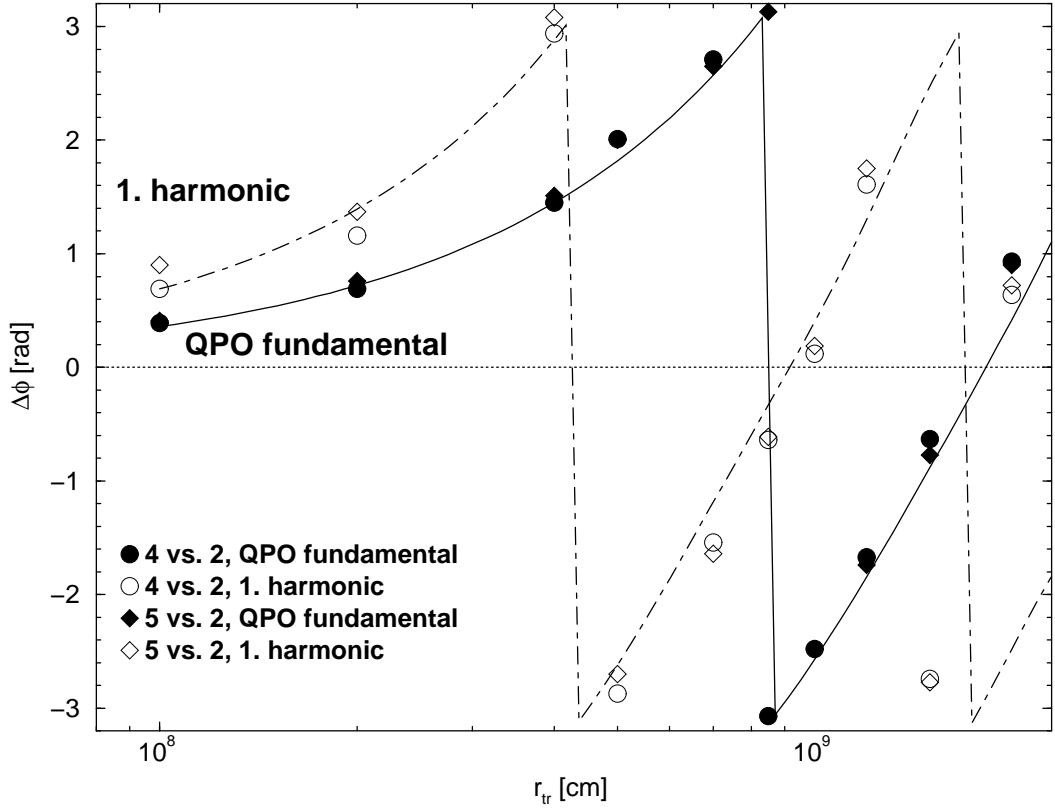}
}
\caption[]{Comparison of the analytical estimates for the phase lags
at the QPO fundamental and 1. harmonic frequencies to the results of
numerical Monte-Carlo simulations. Solid curve: analytical approximation 
for the lag at the QPO fundamental frequency; dot-dashed curve: approximation
for the lag at the 1. harmonic; filled symbols: phase lags at the QPO
fundamental frequency resulting from simulations; open symbols: phase
lags at the 1. harmonic from simulations. Parameters: $f_{\rm QPO} = 10$~Hz,
$\tau_{\rm T}^{ADAF} = 2$, $kT_{\rm D} (r_{\rm tr}) = 0.5$~keV. The energy channels are
[1] 0.1 -- 2.0~keV; [2] 2.0 -- 5.4~keV; [3] 5.4 -- 15~keV; [4] 15 -- 40~keV;
[5] 40 -- 100~keV.}
\label{analytic_comp}
\end{figure}

In order to test our Monte Carlo code against the analytical estimates
derived in the previous section, we did a series of simulations in which 
the accretion disk temperature at the equilibrium transition radius and
the QPO frequency were artificially chosen artificially constant among 
different simulations, and at a value such that the emission in the 
2 -- 5 keV reference channel was always dominated by direct disk 
emission, while the highest two energy channels (15 -- 40~keV and 
40 -- 100~keV, respectively) were dominated by Compton-upscattered 
radiation from the corona. In that case, Eqs. (\ref{tan_Q_le}) -- 
(\ref{tan_H_he}) with constant values of $\alpha$ and 
$\beta/r_{\rm tr}$ may be used in order to approximate the expected 
phase lags. In Fig. \ref{analytic_comp}, the phase lags at the QPO 
fundamental and first harmonic frequencies as measured in our 
simulations are compared to these analytic estimates. The figure
demonstrates that the two approaches are in excellent agreement,
indicating that our numerical procedure reliably reproduces the
time-dependent radiation transport properties of the model system. 

\section{\label{grs1915}The 0.5 -- 10 Hz QPOs in GRS~1915+105}

As briefly mentioned in the introduction, the variable-frequency QPO
at 0.5 -- 10~Hz observed in the low-hard state of GRS~1915+105 exhibits
a very peculiar phase lag behavior (\cite{reig00,lin00}): If the source
has a rather hard (photon index $\gamma \lesssim 2.7$) photon spectrum
and low X-ray flux, the QPO frequency is low $f_{\rm QPO} \lesssim 2$~Hz,
and the phase lags associated with both the QPO fundamental and the
first harmonic frequencies are positive. As the X-ray flux increases
and the photon spectrum becomes softer, the QPO centroid frequency
increases, and the phase lag at the QPO fundamental frequency decreases 
and changes sign at $f_{\rm QPO} \sim 2.5$~Hz. At the same time, the phase
lag associated with the first harmonic frequency increases. As the 
source becomes more X-ray luminous and the spectrum becomes softer, 
the QPO frequency increases up to $\sim 10$~Hz, and the phase lag 
associated with the QPO fundamental continues to decrease, while the 
phase lag at the first harmonic frequency remains positive and does 
not show any obvious correlation with spectral parameters or the QPO 
frequency. Spectral fits to GRS~1915+105 with a disk blackbody + 
power-law model to different spectra states along this sequence 
revealed that the inner disk temperature increases from $\sim 0.7$~keV 
to $\sim 1.5$~keV as the QPO frequency increases from 0.5 to 10~Hz 
(\cite{mmr99}). 

In order to model this behavior, we parametrize the disk temperature
at the transition radius, $T_{\rm D} (r_{\rm tr})$, and the QPO
frequency as power-laws in $r_{\rm tr}$. Assuming $T_{\rm D} (r_{\rm tr})
\propto r_{\rm tr}^{-3/4}$, a simple power-law relation spanning the
observed ranges of $0.7 \, {\rm keV} \le kT_{\rm D}  \le 1.5 \, {\rm keV}$
and $0.5 \, {\rm Hz} \le f_{\rm QPO} \le 10 \, {\rm Hz}$ (\cite{mmr99})
yields a scaling $f_{\rm QPO} \propto T_{\rm D}^4 \propto r_{\rm tr}^{-3}$. 
This scaling may indicate that the transition radius oscillations 
are related to a modulation of the disk evaporation process, 
responsible for the transition to the inner ADAF, by a secular 
instability. The frequency of such modulations might be proportional 
to the inverse of the evaporation time scale, $\tau_{\rm evap}^{-1}$, 
which is expected to be proportional to the disk surface flux, 
$dL/dA \propto T_{\rm D}^4 \propto r_{\rm tr}^{-3}$. Thus, the 
assumed scaling laws of the disk temperature and QPO frequency 
with the transition radius are physically plausible. 

For the high-QPO-frequency end of the sequence mentioned above,
we choose a transition radius of $r_{\rm tr} = 6 \cdot 10^8$~cm,
corresponding to $\sim 700 \, R_s$ for a $3 \, M_{\odot}$ black
hole or the theoretical limit of $340 \, R_s$ found by Meyer et 
al. (\markcite{meyer00}2000) for a $\sim 6 \, M_{\odot}$ black hole
(unfortunately, the mass of the central object in GRS~1915+105
is not known due to the lack of an optical counterpart). The
QPO frequency (i.e. the transition-radius oscillation frequency)
is 10~Hz, and the disk temperature at the transition radius is
1.5~keV. Assuming $M = 6 \, M_{\odot}$ and $r_{tr} = 340 \, R_s$,
this is in agreement with the cool, gas-pressure dominated disk 
model for $\alpha^{-1/5} \, (L / 0.057 L_{\rm Edd})^{3/10} \sim 1$,
where $\alpha$ is the viscosity parameter.

Using the Monte-Carlo simulations described in the previous section,
we calculate the photon-energy dependent light curves, Fourier transform
them, and calculate the Fourier-frequency dependent phase lags from the 
cross-correlation functions. For comparison with the results of
Lin et al. (\markcite{lin00}2000) we sample the light curves in
the energy channels 0.1 -- 3.3~keV [1], 3.3 -- 5.8~keV [2], 5.8
-- 13~keV [3], 13 -- 41~keV [4], and 41 -- 100~keV [5], and focus
on the phase lags between channels [4] and [2]. We perform a series
of 6 simulations, each with different values of $r_{\rm tr}$, implying
different QPO frequencies and disk temperatures according to the scaling 
laws quoted above. 

\begin{figure}
\rotate[r]{
\epsfysize=12cm
\epsffile[150 0 550 500]{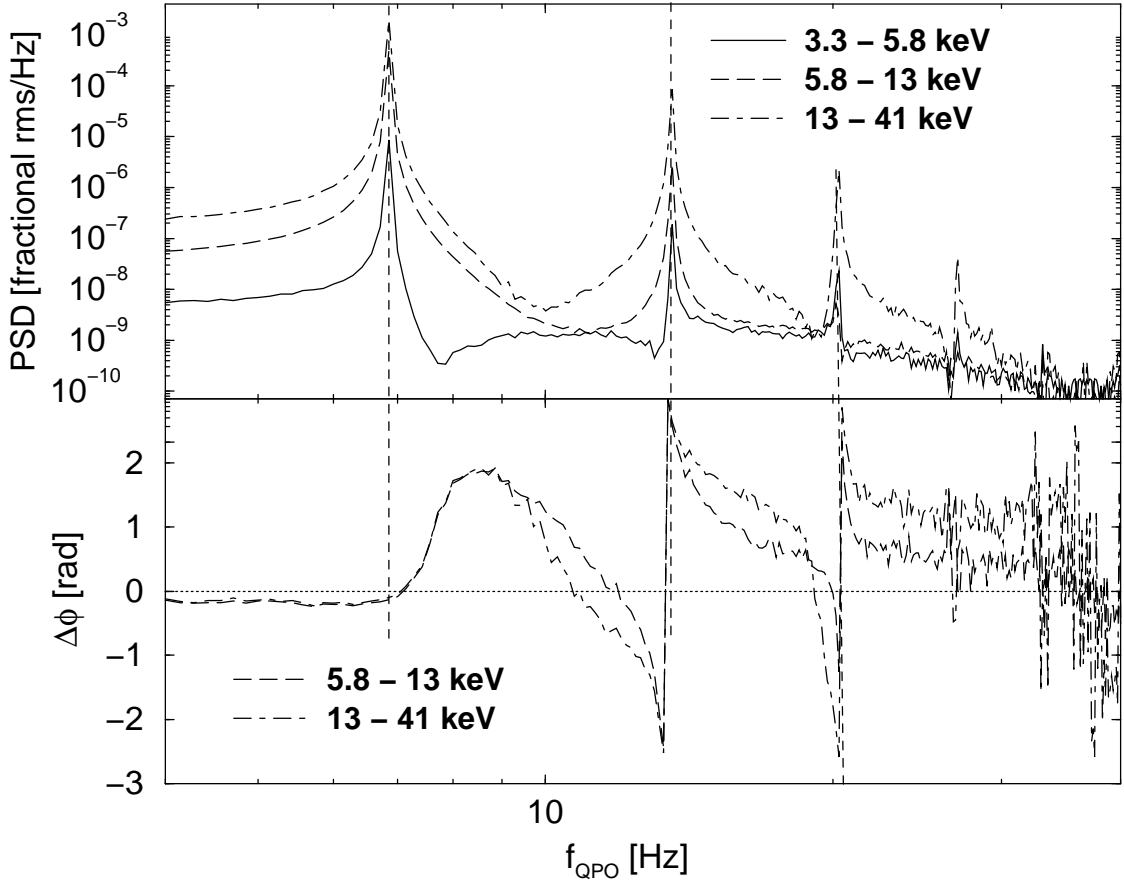}
}
\caption[]{Power spectra (top panel) and phase lags with respect to
the 3.3 -- 5.8~keV channel (bottom panel) of our model simulations at 
a high QPO frequency of $f_{\rm QPO} = 6.7$~Hz, with $r_{\rm tr} = 7 
\times 10^8$~cm. Disk temperature: $kT_{\rm D} (r_{\rm tr}) =
1.4$~keV. Dashed vertical lines indicate the location of the QPO
and its harmonics. The phase lag at the QPO fundamental frequency is
negative, while the lag at the first harmonic frequency is positive. 
For the second harmonic, a negative lag is predicted.}
\label{7Hz}
\end{figure}

\begin{figure}
\rotate[r]{
\epsfysize=12cm
\epsffile[150 0 550 500]{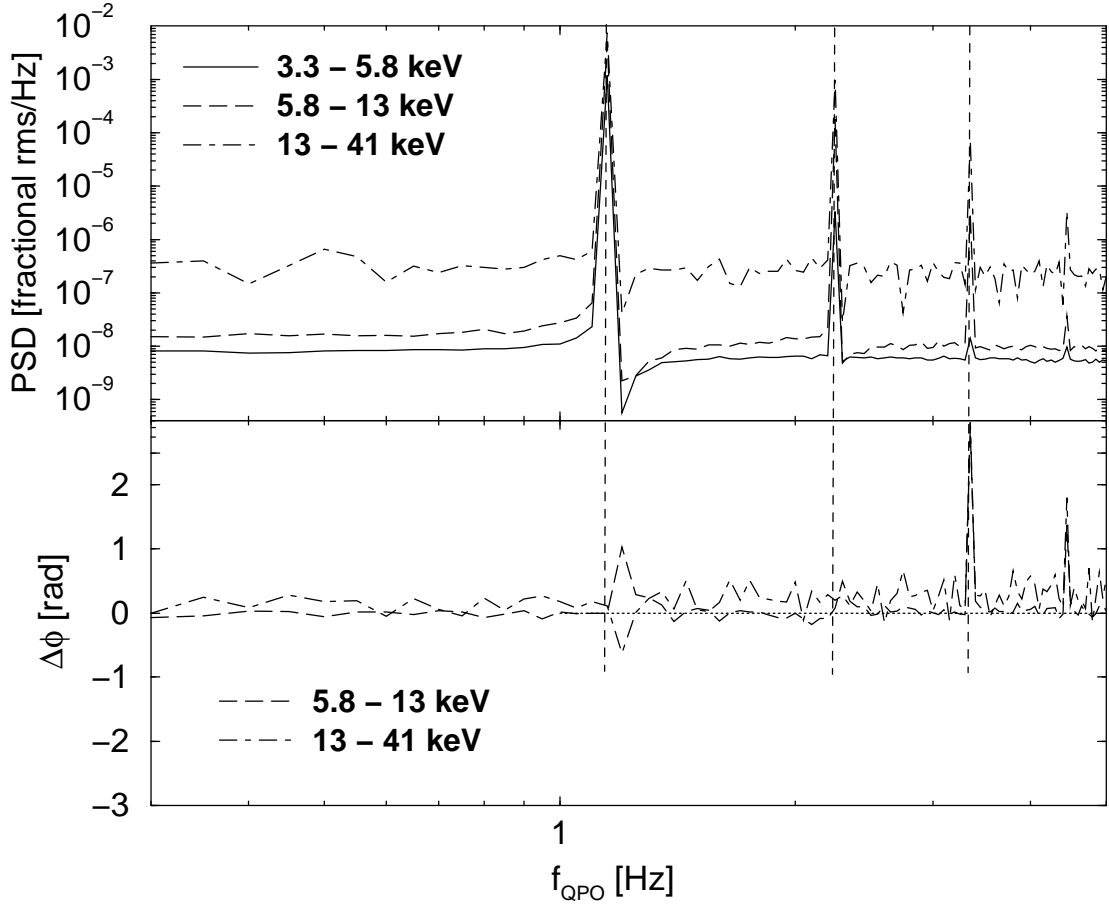}
}
\caption[]{Power spectra (top panel) and phase lags with respect to
the 3.3 -- 5.8~keV channel (bottom panel) of our model simulations 
at a low QPO frequency of $f_{\rm QPO} = 1.1$~Hz, with $r_{\rm tr} = 
1.3 \times 10^9$~cm. Disk temperature: $kT_{\rm D} (r_{\rm tr}) =
0.8$~keV. Dashed vertical lines indicate the location of the QPO
and its harmonics. The phase lags at both the QPO fundamental and the first 
harmonic frequencies are positive for the 13 - 41~keV channel. This 
simulation predicts a positive phase lag also at the second harmonic.}
\label{1Hz}
\end{figure}

Two representative examples of the resulting power and phase lag spectra
are shown in Figs. \ref{7Hz} and \ref{1Hz} for a high-frequency (6.7~Hz)
case with alternating phase lags and a low-frequency (1.1~Hz) case with 
positive phase lags at both fundamental and first harmonic frequencies, 
respectively. 

\begin{figure}
\rotate[r]{
\epsfysize=12cm
\epsffile[150 0 550 500]{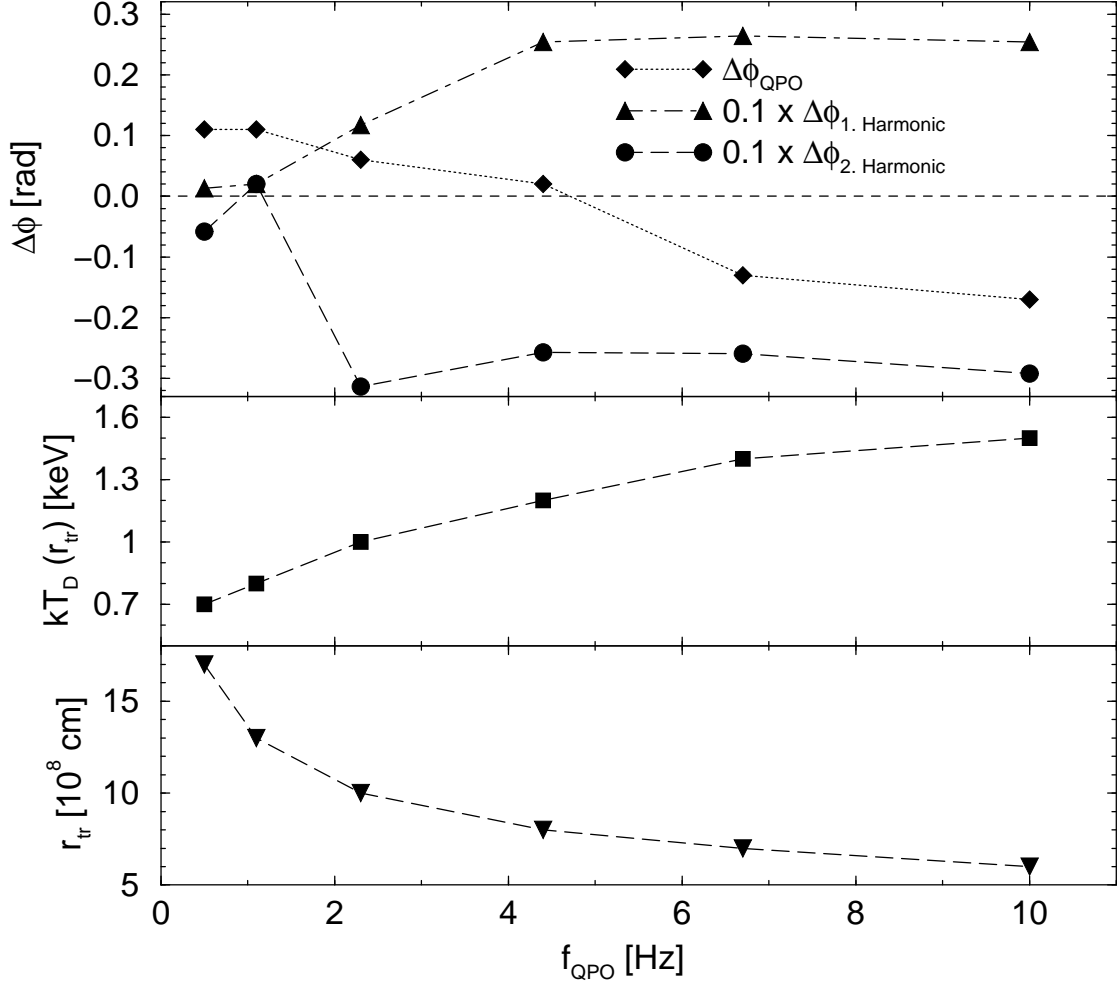}
}
\caption[]{Phase lags between the 13 -- 41~keV and the 3.3 -- 5.8~keV 
energy channel at the QPO fundamental and 1. harmonic frequencies 
(upper panel) resulting from our Monte-Carlo simulations, with inner 
disk temperature as a function of QPO frequency (middle panel) 
determined from observations of GRS~1915+105 and disk/ADAF transition 
radius (lower panel) determined through $T_{\rm D} (r_{\rm tr}) \propto 
r_{\rm tr}^{-3/4}$. The simulations properly reproduce the changing
sign of the phase lag as the QPO frequency increases, and alternating
phase lags for large QPO frequencies, as observed in GRS~1915+105.}
\label{fit1915}
\end{figure}

The results of the complete series of simulations are listed in 
Table \ref{tab1915} and illustrated in Fig. \ref{fit1915}. From 
Fig. \ref{fit1915} we see that this model reproduces the key features 
of the peculiar QPO phase lag behavior observed in GRS~1915+105, in 
particular the change of sign of the phase lag associated with the
QPO as the QPO centroid frequency increases, and the alternating
phase lags between QPO fundamental and first harmonic at high
QPO frequencies. We point out that the measured values of the hard
lags at the first harmonic frequency for $f_{\rm QPO} \gtrsim 2$~Hz
are significantly smaller than predicted by our model. However, this
is expected because, for realistic signal-to-noise ratios, the 
(random-phase) Poisson noise contaminating the actual data always 
suppresses high phase lag values, but leaves small values of 
$\vert\Delta\phi\vert$ almost unaffected (Zhang 2000, private 
communication). 

Table \ref{tab1915} and Fig. \ref{fit1915} also show the predicted
phase lags at the second harmonic of the QPO, which has not been
observed yet. We expect that, if a second harmonic is found in any
future or archived observation of GRS~1915+105, it should be associated
with a negative phase lag, at least for $f_{\rm QPO} \gtrsim 2$~Hz.

We need to point out that the more complicated case discussed in this
section is not directly comparable to the test cases illustrated in
Fig. \ref{analytic_comp} since we are now varying the accretion disk
temperature and the QPO frequency when varying the transition radius.
This also changes the parameters $\alpha$ and $\beta$ entering the
analytical estimates as a function of $r_{\rm tr}$ in a non-trivial
way.

\section{\label{summary}Summary and conclusions}

We have investigated the time-dependent X-ray emission from an
oscillating two-phase accretion flow, consisting of an outer, cool,
optically thin accretion disk, and an inner ADAF. Based on this
model, we are proposing an explanation for the peculiar phase lag
behavior associated with the variable-frequency 0.5 -- 10~Hz QPOs 
observed in GRS~1915+105. In particular, the changing sign of the
phase lag at the QPO fundamental frequency as the QPO frequency 
increases, and the alternating phase lags between QPO fundamental
and first harmonic frequencies in the case of a high QPO frequency
are well reproduced by this model. The relation between QPO frequency, 
transition radius and inner disk radius can be interpreted physically, 
if the QPO is triggered by a modulation of the transition radius by a 
secular instability affecting the disk evaporation responsible for 
the disk/ADAF transition. Based on our results for GRS~1915+105,
we predict that, if a second harmonic to the variable-frequency
QPO at $f_{\rm QPO} \gtrsim 2$~Hz is detected, it should be 
associated with a negative phase lag.

We found that oscillations at the transition radius can naturally 
lead to apparent soft lags associated with the QPO frequency, if 
the photon diffusion time related to the production of hard X-rays 
is longer than half a QPO period. We point out that it is very
unlikely that the same mechanism is responsible for the alternating
phase lags observed in the 67~mHz QPOs and its harmonics since this
would require a transition radius at $r_{\rm tr} \gtrsim 10^5 \, R_s$. 
This large size of the ADAF is implausible for a non-quiescent state.
The 67~mHz QPO alternating phase lag pattern has been observed in a 
high-flux state of GRS~1915+105, in which we would generally expect 
the disk to extend down to rather small radii. 

One caveat of the model we have proposed here is that it requires
a rather large accretion rate in order to produce the observed
disk luminosity from GRS~1915+105. This, in turn, requires either
that a large fraction of the accreted mass is lost to the collimated
outflow of the radio jets, or that the ADAF has a very low radiative 
efficiency in order not to overproduce the hard power-law emission 
from the hot ADAF. However, since GRS~1915+105 is known to have 
very powerful, superluminal radio jets (\cite{mr94}), it is 
plausible to assume that, indeed, a significant fraction of 
the mass accreted through the outer disk is not advected through 
the ADAF, but is powering the radio jets.

\acknowledgements{The work of MB is supported by NASA through
Chandra Postdoctoral Fellowship Award Number PF~9-10007, issued
by the Chandra X-ray Center, which is operated by the Smithsonian
Astrophysical Observatory for and on behalf of NASA under contract
NAS~8-39073.}

\eject

\begin{deluxetable}{cccccc}

\tablewidth{14cm}
\tablecaption{Results of our Monte-Carlo simulations to reproduce
the phase lag behavior of GRS~1915+105. The phase lags $\Delta\phi$
[rad] are the hard lags between channels [4] (13 -- 41~keV) and [2] (3.3
-- 5.8~keV).}
\tablehead{
\colhead{$r_{\rm tr}$ [cm]} & 
\colhead{$f_{\rm QPO}$ [Hz]} & 
\colhead{$kT_{\rm D}$ [keV]} & 
\colhead{$\Delta\phi_{\rm QPO}$} & 
\colhead{$\Delta\phi_{\rm 1. \, harm.}$} &
\colhead{$\Delta\phi_{\rm 2. \, harm.}$}}
\startdata
$1.7 \times 10^9$ & 0.5 & 0.7 & +0.11 & +0.13 & -0.58 \nl
$1.3 \times 10^9$ & 1.1 & 0.8 & +0.11 & +0.20 & +0.20 \nl
$1.0 \times 10^9$ & 2.3 & 1.0 & +0.06 & +1.71 & -3.13 \nl
$8.0 \times 10^8$ & 4.4 & 1.2 & +0.02 & +2.54 & -2.57 \nl
$7.0 \times 10^8$ & 6.7 & 1.4 & -0.13 & +2.64 & -2.59 \nl
$6.0 \times 10^8$ & 10.0& 1.5 & -0.17 & +2.54 & -2.92 \nl
\enddata
\label{tab1915}
\end{deluxetable}

\eject

\eject

\eject

\eject

\end{document}